\documentstyle[12pt,epsf]{article}

\if@twoside
 \else
 \fi

 \parskip \medskipamount
\begin{document}
\title{Chern-Simons number diffusion in (1+1)-dimensional Higgs theory}

\author{Ph. de Forcrand and A. Krasnitz
	\thanks{Address after 1 September 1994: Niels Bohr Institute,
	Blegdamsvej 17, DK-2100 Copenhagen, Denmark.}
\\
        Interdisziplin\"ares Projektzentrum f\"ur Supercomputing\\
	Eidgen\"ossische Technische Hochschule\\
	CH-8092 Z\"urich, Switzerland\\
        \\
        R. Potting\\
        Universidade do Algarve\\
        Unidade de Ci\^encias Exactas e Humanas\\
        Campus de Gambelas, 8000 Faro, Portugal}
\date{July 1994}
\maketitle

\begin{abstract}
We study the Chern-Simons number diffusion rate in the (1+1)-dimensional lattice
Abelian Higgs model at temperatures much higher than, as well as comparable
to, the sphaleron energy.  It is found that in the high-temperature limit
the rate is likely to grow as power 2/3 of the temperature. In the
intermediate-temperature regime, our numerical simulations show that very
weak temperature dependence of the rate, found in previous work, persists at
smaller lattice spacings. We discuss possibilities of relating the observed
behavior of the rate to static finite-temperature properties of the model.
\end{abstract}
\vspace{-20cm}
\begin{flushright}
IPS Research Report No. 94-09
\end{flushright}
\newpage
The Abelian Higgs model (AHM)  in 1+1 dimensions is one of the simplest yet 
nontrivial field-theoretic analogs of the Standard Model (SM) as far as fermion
number nonconservation at high temperature is concerned. Clearly, there is
little hope of quantitatively understanding this phenomenon in SM if a 
similar goal cannot be achieved for AHM, which lends itself much more easily to
both analytical and numerical studies. This explains the considerable amount of
work done on this aspect of AHM in recent years 
\cite{GRS,BTs,BdFH,RHB,ST,GST,IM}.

Prior to the present work, the status of knowledge on thermally induced
fermion-number violation in AHM has been as follows. Much like in SM,
the dynamics of the Chern-Simons number $N_{\rm CS}$ (and thus of the fermion
number) in AHM can reliably be studied in the classical approximation at 
temperatures well above the gauge- ($m_W$) and Higgs- ($m_H$) boson mass 
scales. The relative simplicity of
AHM allows to give an accurate estimate of the $N_{\rm CS}$ diffusion rate at
temperatures $T$ low compared to the energy $E_{\rm sph}$ of a sphaleron 
saddle-point configuration separating vacua labeled by consecutive integer 
values of $N_{\rm CS}$. Not surprisingly, the rate $\Gamma$ exhibits 
thermal-activation behavior \cite{BTs}
\begin{equation}
\Gamma=f(\xi)T^{2/3}\left({E_{\rm sph}\over T}\right)^{7\over 6}
\exp\left(-{E_{\rm sph}\over T}\right),{\ \ \ } m_W,m_H\ll T\ll E_{\rm sph},
\label{BTs}\end{equation}
where $\xi=2m_W^2/m_H^2$. Unlike the SM case, in AHM both the sphaleron energy 
and the $f$ prefactor are 
known analytically. Numerical real-time simulations of AHM at low temperatures,
including those of the current work, indeed agree with (\ref{BTs})
\cite{GRS,RHB,ST}.

Surprising as it may sound, at the opposite end of thermal scale, 
$T\gg E_{\rm sph}$, less was known about the rate in AHM than in SM. In SM, the
only dimensionful quantity relevant at such high temperatures is the temperature
itself, hence, on dimensional grounds, the rate is expected to behave as $T^4$.
In AHM, however, both the gauge and the quartic Higss coupling constants are 
dimensionful, and therefore,
while a power-law behavior was generally believed to be satisfied 
\cite{MW,Dine},
the corresponding power could not be established from dimensional 
considerations. Nevertheless, a simple scaling argument given in the following
indicates that this power is $2/3$. This power law is also consistent with
our numerical results.

As usually is the case, the intermediate-temperature regime ($T$ comparable to
$E_{\rm sph}$) is the least accessible to analytical estimates. Numerical work
\cite{RHB,ST} revealed a surprising feature: at temperatures 5 to 7 times less 
than
$E_{\rm sph}$ the rate practically stops growing and remains roughly constant
until the sphaleron scale of temperatures is reached. Our simulations 
show that this
phenomenon persists at smaller lattice spacings and is therefore not a lattice
artifact.

We shall now elaborate on the points just made,
discussing first the physics and afterwards explaining our 
numerical methods. We begin by giving a
scaling argument in favor of the $T^{2/3}$ behavior of the rate at high 
temperatures. Consider the lattice-regularized classical AHM whose 
temporal-gauge Hamiltonian can, with a suitable choice of units, be written
in a dimensionless form
\cite{BdFH,RHB}:
\begin{equation}
H(v,a)={1\over{2a}}\sum_{j=1}^N[\xi E_{j,j+1}^2 +|\pi_j|^2
+{a^2\over 2}(|\phi_j|^2-v^2)^2
+|\phi_{j+1}-e^{iaA_{j,j+1}}\phi_j|^2],
\label{hxy}
\end{equation}
where $\pi_j$ and $E_{j,j+1}$ are canonically conjugate momenta of
the complex scalar (Higgs) field $\phi_j$ and the spatial component $A_{j,j+1}$
of the gauge field, residing respectively on sites and links of an
$N$-site chain. The lattice spacing is $a$. Periodic boundary conditions are
assumed. The reader can verify by direct substitution that if $\phi_j(t)$, 
$\pi_j(t)$, $A_{j,j+1}(t)$, $E_{j,j+1}(t)$ are solutions at time $t$
of the equations of 
motion following from $H(v,a)$ with $H(v,a)={\cal E}$, then $c\phi_j(ct)$,
$c\pi_j(ct)$, $cA_{j,j+1}(ct)$, $cE_{j,j+1}(ct)$ play the same role 
at time $t$ for
$H(cv,a/c)$ with $H(cv,a/c)=c^3{\cal E}$. Because of this property,
scaling laws can be derived for both static and dynamical observables in AHM.
Consider in particular the Chern-Simons number defined as
\begin{equation}
N_{\rm CS}(t)={a\over{2\pi}}\sum_{j=1}^N A_{j,j+1}(t).
\end{equation}
Its diffusion rate is
\begin{equation}
\Gamma(a,v,T)={\rm lim}_{N,t\rightarrow\infty}{1\over{aNt}}
\langle\left(N_{\rm CS}(t)-N_{\rm CS}(0)\right)^2\rangle_T. \label{ratedef}
\end{equation}
By $\langle\rangle_T$ we mean integration over all the 
initial values of dynamical variables \cite{aper} with the Boltzmann weight
\begin{equation}
\exp\left(-{{H(v,a)}\over T}\right)\prod_{j=1}^N \delta\left(C_j(a)\right),
\end{equation}
where the delta-function factors impose a set of Gauss' constraints
\begin{equation}
C_j(a)\equiv {1\over a}(E_{j,j+1}-E_{j-1,j})-{\rm Im}(\pi_j\phi_j)=0.
\end{equation}
Multiplying all the integration variables ({\it i.e.}, all the initial values
of fields and momenta) in (\ref{ratedef}) by $c$ and using the scaling
property of $H(v,a)$ under this transformation we obtain
a simple scaling rule for the rate:
\begin{figure}[t]
\epsfxsize 10.cm
\epsfbox{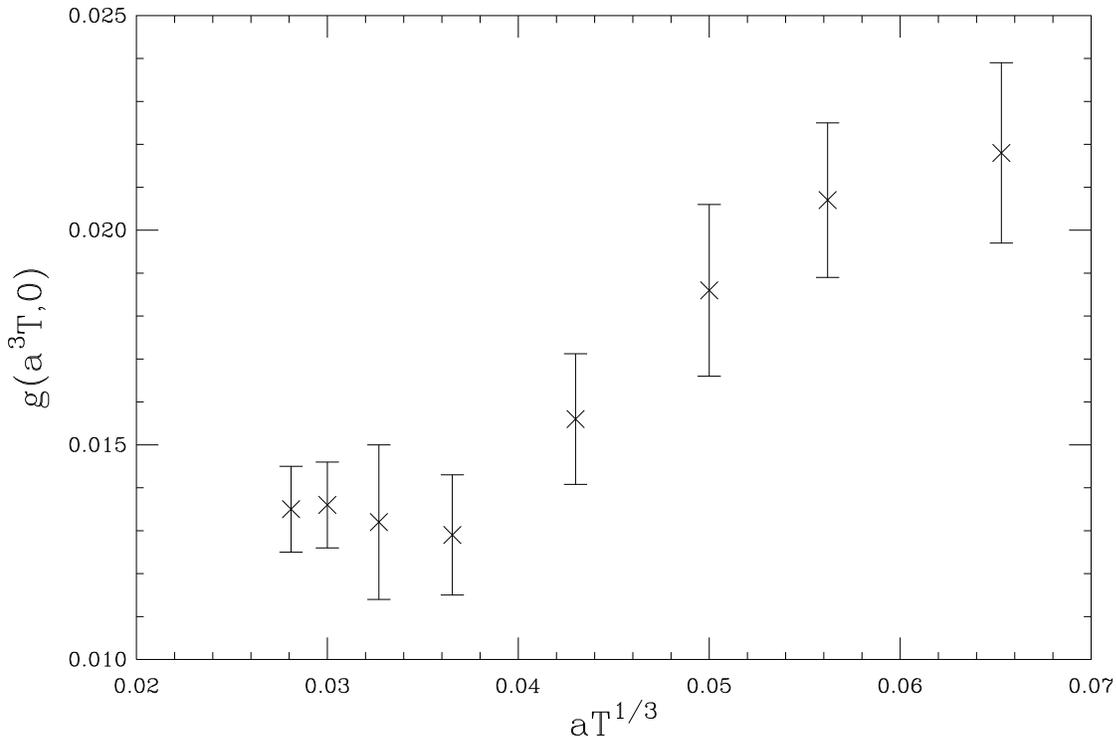}
\caption{$g(a^3T,0)$ measured at small values of $aT^{1/3}$.}
\label{gfun}
\end{figure}
\begin{equation}
\Gamma\left({a\over c},cv,c^3T\right)=c^2\Gamma(a,v,T). \label{scrule}
\end{equation}
Differentiating with respect to $c$ at $c=1$ gives a differential equation
for $\Gamma$ as a function of $a,v,T$:
\begin{equation}
(v\partial_v+3T\partial_T-a\partial_a-2)\Gamma=0,
\end{equation}
whose most general solution is
\begin{equation}
\Gamma=T^{2\over 3} g\left(a^3T,{v^3\over T}\right),
\label{ratescal}\end{equation}
where $g(x,y)$ is an arbitrary function of its two arguments. In particular,
since $E_{\rm sph}\propto v^3$, (\ref{BTs}), derived for the continuum
theory, shows that $g(0,y)\propto y^{7/6}\exp (-y)$ for $y\gg 1$. As for the 
temperatures far above the sphaleron scale, the continuum limit of the rate,
if it exists, is $g(0,0) T^{2/3}$ as claimed. Both result (\ref{BTs}) and
numerical evidence strongly suggest that the rate does indeed have a 
finite continuum limit for any nonzero $v^3/T$. It is also plausible that the
rate becomes insensitive to the details of the Higgs potential at energy scales
far below $T$, {\it i.e.}, $\Gamma$ becomes independent of $v$ if $v^3\ll T$.
We therefore expect $0<g(0,0)<\infty$.
\begin{figure}[t]
\epsfxsize 10.cm
\epsfbox{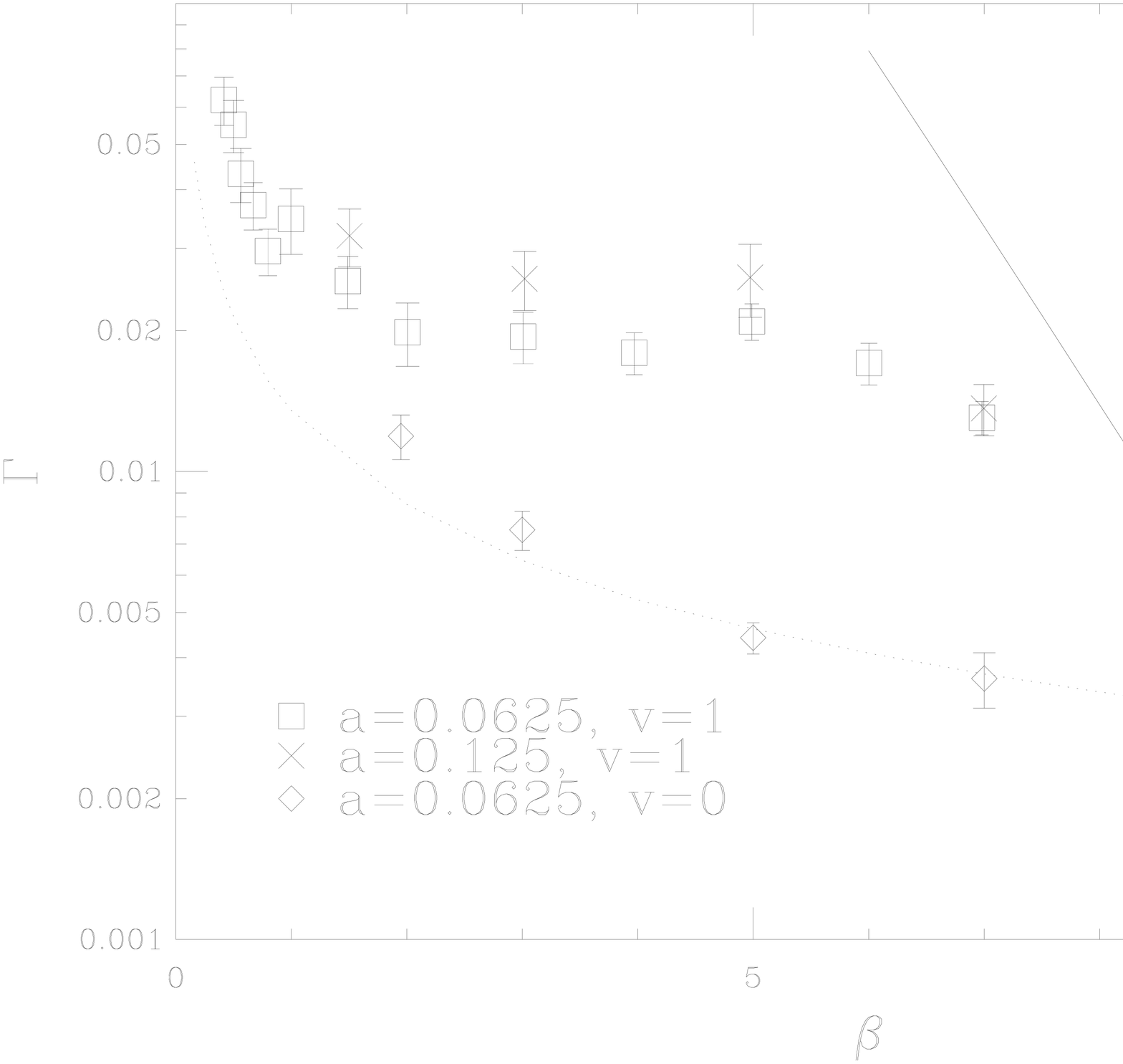}
\caption{Measured rate dependence on the inverse temperature $\beta$. The solid
line is the activation-theory prediction. The dashed line shows the
$T^{2/3}$ law for the high-temperature theory, with $g(0,0)$ estimated at
$\beta=11$, $a=1/16$.}
\label{summary}
\end{figure}

We verified the scaling rule (\ref{scrule}) by a numerical simulation. Since
in this part of the work we are primarily interested in the high-temperature 
regime, we simplify matters by setting $v=0$. The rate measurement then gives
\begin{equation}
{{\Gamma(a=1/2,T=1)}\over{\Gamma(a=1,T=1/8)}}=4.\pm 0.4,
\end{equation}
in agreement with (\ref{scrule}). Next, we performed a series of rate 
measurements closer to the continuum limit of the $v=0$ theory in order to
estimate $g(0,0)$. The measurements, done at lattice spacings 1/16 and 1/8,
span a range of values of the scaling-invariant parameter $aT^{1/3}$ between
0.028 and 0.065. The results, presented in Figure~\ref{gfun}, indicate that 
$g(x,0)$ approaches a finite value as $x\rightarrow 0$.
\begin{figure}[t]
\epsfxsize 10.cm
\epsfbox{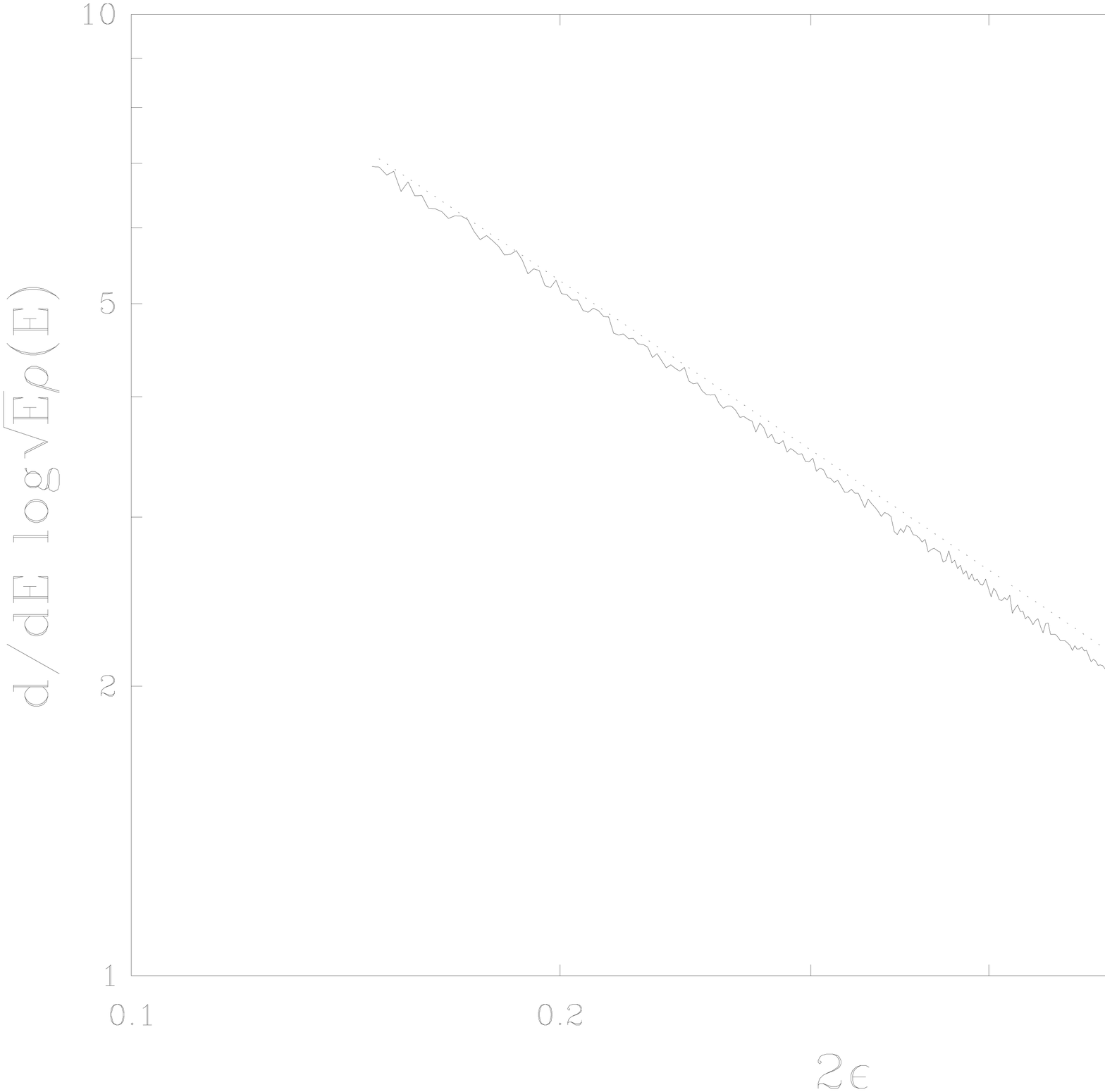}
\caption{Relative variation of the density of states with energy
per degree of freedom
in AHM (solid line) and in a free-field theory (dashed line).}
\label{entropy}
\end{figure}

In this work we are concerned with classical AHM. Note, however, that if the
classical approximation for the rate is valid in quantum AHM at very high
temperatures, so is the $T^{2/3}$ law. Indeed, for consistency of the
classical approximation the classical lattice spacing is chosen inversely
proportional to the temperature \cite{GRS,ST}, yielding $a^3T\rightarrow 0$, 
as required in order for the $T^{2/3}$ law to hold.

Next, we studied the intermediate-temperature properties of the rate.
As in previous work on the subject, we fixed $v=1$, resulting in
$E_{\rm sph}\approx 0.94$. The inverse-temperature range of interest is then
$1<\beta<5$, where an extremely weak temperature dependence of the rate was
found in previous work \cite{RHB,ST}.
By simulating the theory closer to continuum limit ($a=1/16$ compared to
$a\geq 1/8$ in the earlier work), we verified that this feature is not
a lattice artifact: as Figure~\ref{summary} shows, the plateau region becomes 
wider and more pronounced as the lattice spacing shrinks. At still lower 
temperatures, $\beta\geq 7$, the lattice-spacing dependence of the rate
rapidly diminishes. We find a good agreement between our low-temperature rate
measurements and those of \cite{RHB}, performed at larger lattice spacings.
Both earlier and current results approach the asymptotic form (\ref{BTs}) with
decreasing temperature.

Let us summarize the properties of
$N_{\rm CS}$ diffusion rate in AHM as they emerge from this and previous works. 
At low ($T\ll E_{\rm sph}$) the
rate follows the activation-theory prediction (\ref{BTs}) and is
exponentially suppressed by a Boltzmann factor involving the sphaleron barrier
height. At extremely high temperatures ($T\gg E_{\rm sph}$) the rate, if it
has a finite continuum limit, should grow as $T^{2/3}$. Our simulations do
point to a finite continuum limit of the high-temperature rate. The crossover
between the two extreme regimes involves a temperature interval characterized
by a very slow or null rate growth. A complete theory of $N_{\rm CS}$ diffusion
should contain a derivation of the $T^{2/3}$ law at high temperatures and a
determination of $g(0,0)$ from first principles. It also should explain
how corrections to either high- or low-temperature rate lead to the
observed interpolation between the two extremes. In the absence of such a theory
the best we can do is look for clues that might help construct one.
\begin{figure}[t]
\epsfxsize 10.cm 
\epsfbox{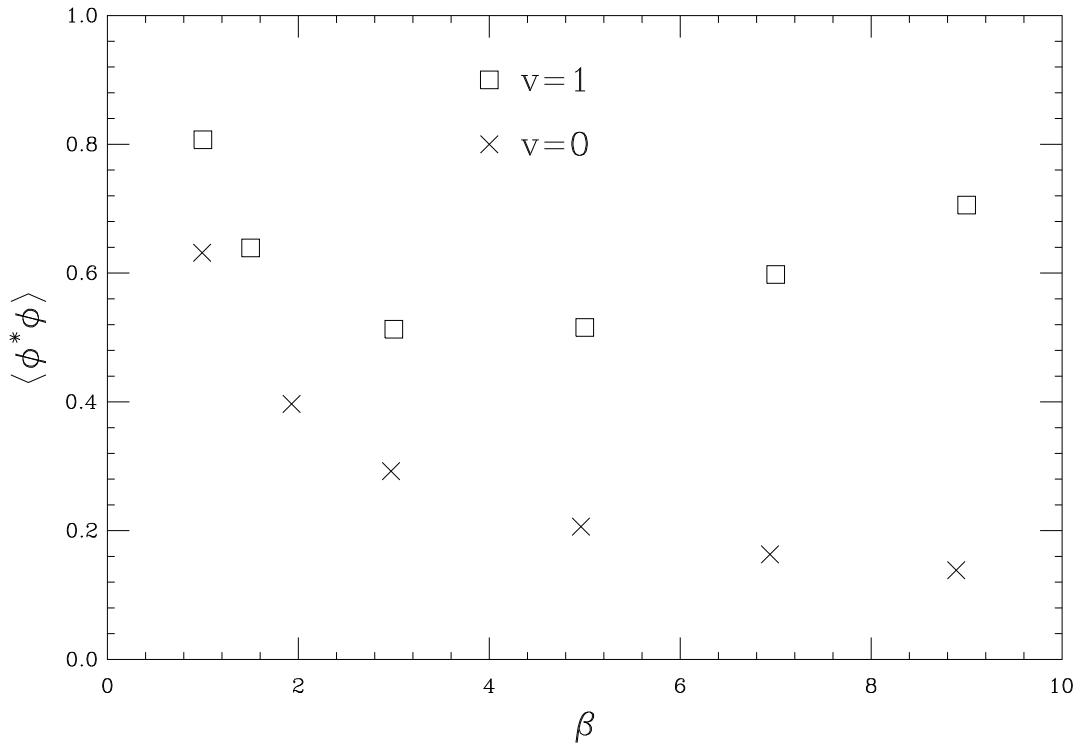}
\caption{Temperature dependence of $\langle|\phi|^2\rangle_T$ in $v=1$
(squares) and $v=0$ (crosses) theories.}
\label{phisquare}
\end{figure}

To this end, we examined 
a possibility that the reasons for the rate suppression
compared to (\ref{BTs}) at intermediate temperatures are entropic in nature
\cite{RHB}.
Namely, the derivation of (\ref{BTs}) is
based on approximating the AHM phase space by that of linear excitations about 
the ground state. Starting at very low temperatures $T$, where it is definitely
valid, this approximation will only hold with increasing $T$ as long as the
free-field relation
\begin{equation}
{d\over{dE(T)}}\log\left(\sqrt{E(T)}\rho(E(T))\right)={1\over{2\epsilon(T)}}
={1\over T} 
\label{dstat} \end{equation}
approximately holds. Here $E(T)$ is the average energy of the system at 
temperature $T$, $\epsilon(T)$ is the average energy per degree of freedom,
and $\rho(E)$ is the corresponding density of states. We determined the 
left-hand side of (\ref{dstat}) using the binning method \cite{bin}.
As Figure~\ref{entropy} shows, the variation of the AHM density of states 
hardly differs from, and even appears somewhat smaller than,
that of a free-field theory, even as the energy per degree of freedom 
(approximately equal to $T/2$ throughout the range of temperatures in question)
approaches
the sphaleron scale. We therefore conclude that 
the observed slowdown of the rate growth is unlikely to be entropy-driven.

In search for other clues we observed that in the intermediate range of 
temperatures ({\it a}) the average squared expectation value 
$\langle\phi^2\rangle_T$ of the scalar field
is close to a minimum (Figures \ref{summary} and \ref{phisquare}); 
({\it b}) the rate in the theory with $v^2=0$
is considerably {\em below} that of $v^2=1$ theory, while
({\it c}) $\langle|\phi|^2\rangle_T$ is considerably greater
in the $v^2=1$ theory than in the $v^2=0$ one. Put together, these pieces of 
empirical evidence suggest that the rate might be closely related to
$\langle|\phi|^2\rangle_T$. Moreover, repeating our scaling argument for
$\langle|\phi|^2\rangle_T$, we obtain an analog of (\ref{ratescal}):
\begin{equation}
\langle|\phi|^2\rangle_T=T^{2\over 3} h\left(a^3T,{v^3\over T}\right),
\label{phiscal}\end{equation}
where, again, we expect $0<h(0,0)<\infty$. As our measurements 
(Table~\ref{phitab}) show,
$h(a^3T,0)=\langle|\phi|^2\rangle_T/T^{2\over 3}$ varies by only about 5\% for
$a^3T$ between $1.9\times 10^{-5}$ and $1.25\times 10^{-4}$. We conclude that,
in all likelihood, $\langle|\phi|^2\rangle_T\propto T^{2/3}$ in the
continuum limit of high-temperature AHM. As a consequence, we also expect
$\Gamma\propto\langle|\phi|^2\rangle_T$ in this limit. It therefore appears
worthwhile to explore the possibility that at intermediate and high temperatures
the time scale for $N_{\rm CS}$ diffusion is primarily set by the value of
$\langle|\phi|^2\rangle_T$. This seems plausible given the
equation of motion for the Chern-Simons number written in the form
\begin{equation}
\ddot N_{\rm CS}={\xi\over{2\pi}}\sum_j\rho_j\rho_{j+1}\sin b_{j,j+1}
={\xi\over{2\pi}} \langle|\phi|^2\rangle_T\sum_j\sin b_{j,j+1}
+{\rm corrections}, \label{effeq}
\end{equation}
where $\alpha_j\equiv{\rm arg}\phi_j$,
$\phi_j=\rho_j\exp(i\alpha_j)$, and $b_j=\alpha_{j+1}-\alpha_j-aA{j,j+1}$.

We conclude by explaining simulation methods used in this work. We chose to
determine $\Gamma$ as given by (\ref{ratedef}) by generating a thermal 
ensemble of initial conditions for lattice AHM. We then let each of these
initial field configurations evolve according to the AHM Hamiltonian equations
of motion. The rate was obtained by averaging $\Gamma$ of (\ref{ratedef})
both over the ensemble of initial conditions and within individual
Hamiltonian trajectories of the system \cite{ST}.
Initial configurations were generated by applying standard Monte-Carlo 
techniques to the theory written in terms of gauge-invariant variables 
\cite{RHB}:
\begin{eqnarray}
H'&=&{a\over 2}\sum_j
\left(\xi\epsilon_{j,j+1}^2
+\left({{\epsilon_{j,j+1}-\epsilon_{j-1,j}}\over{a\rho_j}}\right)^2
+\left({\pi^\rho_j\over a}\right)^2+{2\over a^2}\left(\rho_j^2-
\rho_j\rho_{j+1}\cos b_{j,j+1}\right)\right)\nonumber\\
&&+{a\over 4}\sum_j\left(\rho_j^2-1\right)^2,
\label{hpol}
\end{eqnarray}
\begin{table}
\centerline{\begin{tabular}{|c|rrrrrrr|} \hline
{$a^3T\times 10^5$} & {12.6} & {8.20} & {4.93} & {3.52} & {2.73} & {2.24} &
{1.89} \\ \hline
{$\langle|\phi|^2\rangle_T$} & {$0.615$} & {$0.605$} &
{$0.600$} & {$0.594$} & {$0.595$} & {$0.599$} & {$0.588$} \\
{} & {$\pm 0.005$} & {$ \pm 0.004$} &
{$\pm 0.004$} & {$\pm 0.004$} & {$\pm 0.007$} & {$\pm 0.004$} & {$\pm 0.004$} \\
\hline
\end{tabular}}
\caption{Average squared magnitude of the scalar field as a function of $a^3T$
in the $v=0$ theory.}
\label{phitab}
\end{table}
where $\rho_j$ and $b_{j,j+1}$ are defined as in (\ref{effeq}),
$\epsilon_{j,j+1}=E_{j,j+1}/a$, and $\pi^\rho_j$ is the canonical momentum of
$\rho_j$.  More specifically, we used the Metropolis algorithm to update the 
$\rho$ and $b$ variables and the heat-bath algorithm to update $\epsilon$.
The fourth canonical variable, 
$\pi_\rho$, is normally distributed and required no iterative procedure to
be generated. Note that by choosing to work with gauge-invariant variables we
ensured automatic fulfillment of Gauss' law by initial configurations. 
Having generated an initial configuration, we performed transformation to a
gauge-dependent set of variables as in (\ref{hxy}) under an initial gauge 
condition ${\rm arg}\phi_j=0$, $A_{j,j+1}=-b_{j,j+1}/a$ 
and integrated the Hamiltonian equations of motion in
terms of this latter variable set. We used the fourth-order leapfrog algorithm
\cite{CG}.
In case of AHM and other gauge theories leapfrog algorithms offer, beside their
superior energy-conserving properties, the extra benefit of exact local charge
conservation. For the smallest lattice spacing used, $a=1/16$, we varied the 
time step from 0.02 at low temperatures to 0.005 at high temperatures, keeping
it small enough to suppress any time-step dependence of observables. As a rule,
we set the physical size $Na$ of the system to 50, known to be large enough
to avoid noticeable finite-size effects on the rate and other measured 
observables \cite{RHB}. Finally, the value of $\xi$ was 10 throughout our 
simulations.
\section*{Acknowledgments}
We thank
A.~Bochkarev,
A.~Kovner, J.~Smit, A.~Wipf, W.H.~Tang, E.~Mottola, S.~Habib, and 
W.P.~Petersen for enlightening discussions. PdF and AK
were supported by the Swiss National Science Foundation. RP and AK were 
supported by
JNICT (Portugal) under grant No.~STRDC/C/FAE/1018/93. Numerical simulations
for this work were performed on the Intel Paragon and the Cray-Y/MP
supercomputers at the ETH.


\begin{thebibliography}{9}
\bibitem{aper}Since $H(v,a)$ is a periodic function of the $A$ variables, 
the latter need only be integrated in the range $|aA|\leq\pi$.
\end{thebibliography}
\end{document}